\documentclass[a4paper,12pt]{article}

\usepackage{ragged2e} 
\usepackage{changepage}

\usepackage{siunitx} 
\usepackage{amsmath}
\usepackage{bm} 
\usepackage{amssymb}
\usepackage{wasysym}
\usepackage{xfrac} 

\usepackage{graphicx}
\usepackage{sidecap}
\usepackage{wrapfig}
\usepackage{caption}
\usepackage{subcaption}

\usepackage{verbatim}
\usepackage[english]{babel}

\usepackage{array}
\usepackage{adjustbox}
\usepackage{float}
\usepackage{enumerate}
\usepackage{multirow}
\usepackage{multicol}
\usepackage[table,xcdraw]{xcolor}

\usepackage[utf8]{inputenc}
\usepackage[super,sort&compress]{natbib}
\makeatletter
\renewcommand{\@biblabel}[1]{#1.}
\makeatother
\setcitestyle{comma}
\usepackage{csvsimple}
\usepackage{bbm}
\usepackage{pdflscape}
\usepackage{breqn}
\usepackage{lscape}
\usepackage{amsfonts}
\usepackage{latexsym}
 \usepackage[titletoc]{appendix}  
\usepackage[colorlinks=true, citecolor= black]{hyperref} 
\hypersetup{linkcolor=black} 
\usepackage[all]{hypcap} 
\usepackage{setspace}
\usepackage{enumitem}
\usepackage{authblk}
\usepackage{bm}
\usepackage[T1]{fontenc}
\usepackage{helvet}
\usepackage{listings}
\lstdefinestyle{Rstyle}{
  language=R,
  basicstyle=\ttfamily\scriptsize,
  numbers=left,
  numberstyle=\tiny,
  numbersep=5pt,
  breaklines=true,
  breakatwhitespace=true,
  tabsize=2,
  frame=single,
  showstringspaces=false
}
\usepackage[pagewise]{lineno}

\title{\textbf{Linear mixed modelling of federated data when only the mean, covariance, and sample size are available}}
\author[1]{\small Marie Analiz April Limpoco}
\author[1]{\small Christel Faes}
\author[1,2]{\small Niel Hens}
\affil[1]{Interuniversity Institute for Biostatistics and Statistical Bioinformatics (I-BioStat), Data Science Institute (DSI), Hasselt University, Hasselt, Belgium}
\affil[2]{Centre for Health Economic Research and Modelling Infectious Diseases (CHERMID), Vaccine \& Infectious Disease Institute, Antwerp University, Antwerp, Belgium}

\date{}
\begin{document}
\maketitle


\begin{abstract}
In medical research, individual-level patient data provide invaluable information, but the patients' right to confidentiality remains of utmost priority. This poses a huge challenge when estimating statistical models such as linear mixed models, which is an extension of linear regression models that can account for potential heterogeneity whenever data come from different data providers. Federated learning algorithms tackle this hurdle by estimating parameters without retrieving individual-level data. Instead, iterative communication of parameter estimate updates between the data providers and analyst is required. In this paper, we propose an alternative framework to federated learning algorithms for fitting linear mixed models. Specifically, our approach only requires the mean, covariance, and sample size of multiple covariates from different data providers once. Using the principle of statistical sufficiency within the framework of likelihood as theoretical support, this proposed framework achieves estimates identical to those derived from actual individual-level data. We demonstrate this approach through real data on 15 068 patient records from 70 clinics at the Children's Hospital of Pennsylvania (CHOP). Assuming that each clinic only shares summary statistics once, we model the COVID-19 PCR test cycle threshold as a function of patient information. Simplicity, communication efficiency, and wider scope of implementation in any statistical software distinguish our approach from existing strategies in the literature.
\end{abstract}


{\small \bf Correspondence:} {\small Marie Analiz April Limpoco \\ 
Interuniversity Institute for Biostatistics and Statistical Bioinformatics (I-BioStat),\\ Data Science Institute (DSI), Hasselt University, Hasselt, Belgium \\
Email: liz.limpoco@uhasselt.be}

\section{Introduction}
\noindent Understanding and extracting useful information from data are some of the shared goals between data providers and data analysts. However, both parties must also respect the right to data privacy of individuals from whom the data were collected. This imposes restrictions on how much and which kind of data can be disclosed by the data providers to the data analysts.\citep{european_commission_regulation_2016} Consequently, this adds to the challenge of estimating statistical models. For example, in health research involving patient data from different hospitals, estimating a linear mixed model to account for potential heterogeneity across hospitals requires individual-level patient records. In the interest of data confidentiality, hospitals might be reluctant to share the full data unless the data analyst goes through a series of processes and paperwork, which may take considerable time.\citep{Chiruvella2021-xf}  \\

A possible compromise is to employ federated learning algorithms.\citep{50116} In this setting, only parameter estimate updates and not the individual-level data are sent by data providers to a centralized server to build a global model.\citep{BANABILAH2022103061} Prevalent models for which a federated learning algorithm was developed include linear regression,\citep{doi:10.1080/24725854.2022.2157912} generalized linear mixed models,\citep{Li2022, Yan2023-yn} logistic regression, support vector machine (SVM), K-Means, neural network, Bayesian network, and random forest.\citep{Crowson2022-bt, 10.1145/3501813} To implement this strategy, a network of computer connections that enable iterative communication between the data providers and data analysts has to be set up. In practice, at least in the healthcare context, it is not easy to set up such networks that fulfill the requirements conducive to federated learning.\citep{Wen2023, 9084352, Rieke2020}   \\

Instead of an iterative communication, data providers may be more willing to share summary data once. For a linear regression model, if these summary data contain sufficient statistics, then model estimation is possible even when the individual-level data are unavailable.\citep{doi:10.1080/00031305.2016.1255659} For a linear mixed model involving random effects per data provider, Luo et al\citep{f6c8409ac3ed4b1cb27df3dfa95ac12c} expressed the likelihood in terms of aggregate matrices, which in turn are composed of sufficient statistics. They did this by applying the Woodbury matrix identity and some linear algebra concepts. Since most of the existing functionalities in statistical softwares require the individual observations as input, model estimation using summaries requires the development of new functionalities. Luo et al\citep{f6c8409ac3ed4b1cb27df3dfa95ac12c} developed an R package called \texttt{pda} \citep{R} to implement their proposed method, which they referred to as distributed linear mixed model (DLMM), but its structure is in a distributed model estimation context. Specifically, in practice, data providers must send their aggregate matrices to a central online server.\\

In the meta-analysis setting where the ``data providers" are the relevant studies, there are advantages of using individual participant data (IPD) over aggregate data (AD) especially when the number of studies is small.\citep{Song2020-um, 10.1002/jrsm.1331} However, IPD is seldom available from studies and so constructing a substitute for the unavailable IPD, called pseudo-IPD, was an alternative. One strategy is to require the specification of the joint distribution and correlation structure from which several sets of pseudo-IPD will be simulated to obtain parameter estimates per set.\citep{Song2020-um} These will then be aggregated to arrive at a single estimate per parameter of interest. Since this method involves simulating from an assumed distribution whose parameters are based on the available AD, a point to consider in practice is how many sets should be produced to achieve the desired accuracy.\\ 

Another approach requires only one set of pseudo-IPD to yield estimates exactly equal to the actual IPD estimates.\citep{10.1002/jrsm.1331} In this approach, pseudo-IPD are constructed by first generating random numbers from any distribution and then transforming them such that the mean and standard deviation of the resulting data are exactly equal to those indicated in the studies. Unlike DLMM which also yields exactly the same estimates as the actual IPD estimates, this pseudo-IPD approach can be implemented in any statistical software with existing linear mixed model functionalities. A limitation of both pseudo-IPD methods though, in the context of meta-analysis, is that the covariance structure of the relevant variables is rarely available from studies, thus making it difficult to include more covariates in the model. In particular, their scope only includes studies with two treatments or exposure groups. In the context of federated data, this is not a problem since data providers like hospitals may be willing to supply the covariance structure of the variables, aside from the mean vector. \\

In this paper, we apply the pseudo-IPD meta-analysis framework of Papadimitropoulou et al \citep{10.1002/jrsm.1331} to the federated data setting to estimate a linear mixed model with random effects per data provider when only the mean, covariance, and sample size are made available once. More importantly, we extend it so that multiple covariates can be included, distinguishing our approach from theirs. This principle of using pseudo-data in the context of federated data analysis is novel. To summarize the difference of our approach from those that exist in the literature, we present Table \ref{tab:diff_ourapproach}. Among the existing methods, DLMM and our proposed strategy employ the concept of sufficient statistics and are thus expected to yield theoretically the same results. The major difference lies in how the sufficient statistics are utilized: DLMM uses them directly in the re-expressed form of the likelihood, while we use them to generate pseudo-data before feeding these into the classical form of the likelihood. Hence, our proposed strategy may be considered as an alternative and is not necessarily superior over DLMM; though it has some advantages in terms of generalisability and implementation. The following points distinguish our approach from DLMM:

    \begin{enumerate}
        \item \textbf{Extension to more complex models.} The principle of generating pseudo-data facilitates model building, performing post hoc procedures, and extension to more complex models such as generalized linear models (GLMs) while keeping the communication exchange with the data providers to just once. Although the idea of DLMM has already been extended to GLMs, at least one communication exchange is needed. In addition, in their current framework, only linear models have the option to include random effects.\citep{pdamethods} On the contrary, we have found in our current research work that it is possible to apply a similar principle of pseudo-data generation from one-time shared summary statistics to the other members of the GLM family. While in this paper we focus on linear models, we will discuss the extension towards GLMs in the discussion of the paper.
        
        \item \textbf{Practical implementation.} Our proposed method makes use of the classical form of the likelihood which feeds on individual observations. This is the basis for existing LMM functionalities in various statistical softwares. This means that after the pseudo-data are generated, we can use any of the existing software programs available to fit the mixed model. This is advantageous due to the strength of existing functionalities for linear mixed modelling which can be found not only in R \citep{R} but also in other statistical softwares. This implies that compared to \texttt{pda},\citep{Rpda} current versions of existing software packages (e.g. \texttt{lmer} in \texttt{lme4} \citep{Rlme4}) have been optimized in terms of numerical stability, computational efficiency, optimization algorithms, and bugs fixing.\citep{Rlme4} Furthermore, user support is more available online for possible problems that a user may encounter when using these existing functionalities. Thus, utilizing these existing functionalities may be preferred by practitioners, if possible. 
    \end{enumerate}

    \begin{table}[h!]
            \centering
            \caption{Difference of proposed strategy from existing methods}
            \resizebox{\textwidth}{!}{
            \begin{tabular}{lcccc}
                \hline
                \textbf{Aspect} & \textbf{Our} & \textbf{Federated} & \textbf{DLMM} & \textbf{pseudo-IPD} \\
                 & \textbf{approach} & \textbf{learning} &  & \textbf{in meta-analysis} \\\hline
                Input & \textit{one set of} & parameter estimate & aggregate & \textit{at least one set} \\
                 & pseudo-data & updates until & data & of pseudo-data\\
                 & with \textit{identical} & convergence & matrices & with \textit{similar} \\
                 & sufficient statistics & & & sufficient statistics \\
                 & as actual data & & & as actual data \\
                 & & & & \\
                Number of & multiple & multiple & multiple & one/limited \\
                covariates & & & & \\
                 & & & & \\
                Estimation & use pseudo-data & global model is & likelihood is rewritten & same as our\\
                strategy & as substitute to & estimated iteratively & in terms of aggregate & approach but \\
                 & actual unavailable & from local parameter & matrices instead of & repeated multiple \\
                 & data & estimates of & individual observations & times; estimates\\
                 & & data providers & & aggregated \\
                 & & & & into one\\
                 & & & & \\
                Theoretical estimates & identical to & close to & identical to & close to \\
                produced & estimates from & estimates from & estimates from & estimates from \\
                 & actual data & actual data & actual data & actual data \\
                 & & & & \\
                Communication rounds & once & more than once & once & once\\ 
                between data providers &  & & & \\
                and data analyst & & & & \\
                 & & & & \\
                Infrastructure & none & central server must be & online server where & none\\
                requirements & & connected to data & data providers send & \\
                 & & providers' databases & aggregate matrices & \\
                 & & & & \\
                Software & any statistical & TensorFlow Federated, & R package \texttt{pda} \citep{R} & any statistical\\
                implementation & software with & OpenFL, PySyft & & software with\\
                 & LMM functionality &  & & LMM functionality\\
                \hline
            \end{tabular}}
            \label{tab:diff_ourapproach}
        \end{table}

In the next section (\ref{section:methods}), we present the details of our proposed framework. We then demonstrate it in section \ref{section:example} on deidentified publicly available real data consisting of the results of COVID-19 testing at the Children’s Hospital of Pennsylvania (CHOP) in 2020 which can be found in the R\cite{R} package \texttt{medicaldata}.\cite{Rmedicaldata} Finally, we discuss implications as well as potential research directions in section \ref{section:discussion}, before we close with a conclusion.

\section{Methods}\label{section:methods}
\subsection{Principles of data reduction}\label{suf}
\noindent This section briefly revisits the principles of data reduction, which are thoroughly discussed by Casella and Berger.\citep{CaseBerg:01} We aim to draw insights about dealing with parameter estimation given limited data. We begin with the sufficiency principle, which guarantees that the entire sample need not be available, as inferences about a parameter $\theta$ can be derived from the sufficient statistic $T(\mathbf{X})$, if it exists. In other words, even if the only information known is $T(\mathbf{x})$, inference about the parameter of interest $\theta$ can still be made. In connection with our objective, if the data providers supply sufficient statistics for the model of interest, which in this case is the linear mixed model, then parameter estimation is still possible even without disclosing the individual-level data. \\

Once sufficient statistics are identified, if they exist, parameter estimation can then proceed by directly plugging in the sufficient statistics instead of the individual observations into the log-likelihood of the model. An alternative strategy is to generate a sample $\mathbf{x_2}$ which is different from the original sample $\mathbf{x_1}$, but such that $T(\mathbf{x_1}) = T(\mathbf{x_2})$, and use $\mathbf{x_2}$ as input to the existing functionalities that estimate a linear mixed model. The sufficiency principle will still guarantee that the same conclusion can be drawn even though $\mathbf{x_1} \neq \mathbf{x_2}$. Casella and Berger\cite{CaseBerg:01} explicitly mentioned that the conditional probability distribution given a sufficient statistic $T(\mathbf{x_1})$ can be used to draw a sample $\mathbf{x_2}$ and generate equivalent information about $\theta$. However, they did not discuss how to obtain this probability distribution in practice.  \\

To supplement the sufficiency principle, we use the likelihood principle (Appendix \ref{lik_prin}). Specifically, if we aim to exactly estimate a parameter through its likelihood based on $\mathbf{x_1}$ but only its sufficient statistic $T(\mathbf{x_1})$ is available, we can generate a sample $\mathbf{x_2}$ such that $T(\mathbf{x_1}) = T(\mathbf{x_2})$ and use the likelihood based on $\mathbf{x_2}$ to estimate the parameter of interest. To this end, the distribution that generated $\mathbf{x_2}$ becomes immaterial. In the succeeding sections, we implement this idea. Specifically, we first identify the sufficient statistics and then generate another sample which we refer to as pseudo-data such that the sufficient statistics of the actual and pseudo-data are equal. We start with sufficient statistics for a linear regression model and extend the idea to a linear mixed model. \\

\subsection{Sufficient statistics for a linear regression model}
\noindent The concepts presented in section \ref{suf} were demonstrated by Lee, Brown, and Ryan\citep{doi:10.1080/00031305.2016.1255659} for a linear regression model. In particular, given $n$ observations, an intercept, and $p-1$ predictors, if $\mathbf{X}$ denotes the $n \times p$ design matrix and $\mathbf{y}$ is the $n \times 1$ vector of continuous responses, the linear regression coefficients $\boldsymbol{\beta}$ are estimated as
\begin{align*}
    \hat{\boldsymbol{\beta}} &= (\mathbf{X}^T\mathbf{X})^{-1}\mathbf{X}^T\mathbf{y},
\end{align*}
where knowing the aggregate information $\mathbf{X}^T\mathbf{X}$ and $\mathbf{X}^T\mathbf{y}$ instead of the individual-level data $\mathbf{X}$ and $\mathbf{y}$ will still yield exactly the same regression coefficient estimates $\hat{\boldsymbol{\beta}}$. We expand on this and explicitly show that the sample mean, sample covariance matrix, and sample size are sufficient to produce the same parameter estimates as with using the individual-level data. In addition to the work of Lee, Brown, and Ryan\citep{doi:10.1080/00031305.2016.1255659}, we include estimation of the variance
\begin{align*}
    \hat{\sigma}^2_{MLE} &= \frac{1}{n}\sum_{i=1}^n (y_i - \mathbf{x}_i \hat{\boldsymbol{\beta}})^2, \text{ or} \\
    \hat{\sigma}^2_{OLS} &= \frac{1}{n-p}\sum_{i=1}^n (y_i - \mathbf{x}_i \hat{\boldsymbol{\beta}})^2.
\end{align*} 
For more specific details regarding the derivations, see Appendix \ref{suf_lin_reg}. 

We begin by looking at the log-likelihood
\begin{align*}
    l(\boldsymbol{\beta}, \sigma^2; \mathbf{y}, \mathbf{X}) &= -\frac{n}{2}\text{ln}(2\pi) -\frac{n}{2}\text{ln}(\sigma^2) -\frac{1}{2\sigma^2}\sum_{i=1}^n(y_i-\mathbf{x}_i^T\boldsymbol{\beta})^2,
\end{align*}
where $\mathbf{x}_i$ is a vector representing the $i$th row in the design matrix $\mathbf{X}$. We see here that information from the sample is required only in the last term. Moreover, the sum of squares of this term can be expressed as  
\begin{align*}
    \sum_{i=1}^n(y_i-\mathbf{x}_i^T\boldsymbol{\beta})^2 &= \sum_{i=1}^ny_i^2 - 2\sum_{i=1}^ny_i\mathbf{x}_i^T\boldsymbol{\beta} + \sum_{i=1}^n\boldsymbol{\beta}^T\mathbf{x}_i\mathbf{x}_i^T\boldsymbol{\beta}.
\end{align*}
From this, we find that knowing $n$, $\sum_{i=1}^ny_i^2$, $\sum_{i=1}^ny_i\mathbf{x}_i^T$, and $\sum_{i=1}^n\mathbf{x}_i\mathbf{x}_i^T$ is sufficient to construct the log-likelihood and estimate the parameters, even in the absence of individual-level data. In particular, $\sum_{i=1}^ny_i\mathbf{x}_i^T$ and $\sum_{i=1}^n\mathbf{x}_i\mathbf{x}_i^T$ are sufficient to estimate the coefficients $\boldsymbol{\beta}$ while the variance $\sigma^2$ also requires $\sum_{i=1}^ny_i^2$ in addition to the other two. Furthermore, these values can be obtained from the vector of sample means and sample covariance matrix of the response variable and the predictors. Specifically, performing some algebraic manipulations will show that $\sum_{i=1}^ny_i^2$ can be derived from the sample variance $s^2_{\mathbf{y}}$, the sample mean $\bar{\mathbf{y}}$, and the sample size $n$
\begin{align*}
    \sum_{i=1}^ny_i^2 &= s^2_{\mathbf{y}}(n-1) + n\bar{\mathbf{y}}^2.
\end{align*}
For $\sum_{i=1}^ny_i\mathbf{x}_i^T$, we note that it is a $1\times p$ matrix
\begin{align*}
    \begin{bmatrix}
        \sum_{i=1}^ny_i & \sum_{i=1}^ny_ix_{i1} & \sum_{i=1}^ny_ix_{i2} & ... & \sum_{i=1}^ny_ix_{ij} & ... & \sum_{i=1}^ny_ix_{i(p-1)} 
    \end{bmatrix},
\end{align*}
such that the first element can be obtained from the sample mean $\bar{\mathbf{y}}$ while the rest of the elements needs the sample covariance between $\mathbf{y}$ and each of the predictors ($s_{\mathbf{yx}_j}$). Lastly, since $\sum_{i=1}^n\mathbf{x}_i\mathbf{x}_i^T$is a $p\times p$ matrix
\begin{align*}
    \begin{bmatrix}
        n & \sum_i{x_{i1}} & \hdots & \sum_i{x_{i(p-1)}} \\
        \sum_i{x_{i1}} & \sum_i{x_{i1}^2} & \hdots & \sum_i{x_{i1}x_{i(p-1)}}\\
        \vdots & \vdots & \ddots & \vdots\\
        \sum_i{x_{i(p-1)}} & \sum_i{x_{i(p-1)}x_{i1}} & \hdots & \sum_i{x_{i(p-1)}^2} 
    \end{bmatrix},
\end{align*}
performing similar derivations as above reveals that computing $\sum_{i=1}^n\mathbf{x}_i\mathbf{x}_i^T$ only requires the sample mean ($\bar{\mathbf{x}}_j$), sample variance ($s^2_{\mathbf{x}_j}$), and sample covariances among predictors $j$ and $k$ ($s_{\mathbf{x}_j\mathbf{x}_k}$).

\subsection{Sufficient statistics for a linear mixed model}

\noindent A more realistic assumption when handling federated data is that the observations from the same data provider are more similar than observations from different sources, violating the independence assumption of a linear regression model. To account for this, a linear mixed model is more appropriate. Assuming that there are $m$ data providers, let $y_{hi}$ be the continuous response of individual $i$ from data provider $h$; $\mathbf{x}_{hi}$ be a $p$-dimensional vector consisting of an intercept and $p-1$ predictors; $\boldsymbol{\beta}$ be the p-dimensional vector of fixed effects; $\mathbf{z}_{hi}$ be the $q$-dimensional vector corresponding to the $q$ random effects; ${\mathbf{u}}_h$ be the $q$-dimensional random effects vector, which represents the deviation of data provider $h$ from the overall pattern; and $\epsilon_{hi} \sim N(0, \sigma^2)$ be the random error. For a linear mixed model with random effects for each data provider, the model structure will be
\begin{align*}
    y_{hi} &= \mathbf{x}_{hi}^T\boldsymbol{\beta} + {\mathbf{z}}_{hi}^T{\mathbf{u}}_h + \epsilon_{hi}.
\end{align*}
For a model with a random intercept and slope, $q=2$, $\mathbf{z}_{hi} = [1, z_{hi}]$ and $\mathbf{u}_h \sim N(\boldsymbol{0}, \mathbf{G})$ where $\mathbf{G}$ is the $2 \times 2$ random effects covariance matrix. To estimate parameters, the marginal log-likelihood used is
\begin{align*}
    l(\boldsymbol{\beta}, \sigma^2, \mathbf{G}; \mathbf{y}, \mathbf{X}) &= -\frac{1}{2}\sum_{h=1}^m\{\text{log}|\boldsymbol{\Sigma}_{h}| + (\mathbf{y}_h - \mathbf{X}_h\boldsymbol{\beta})^T\boldsymbol{\Sigma}_{h}^{-1}(\mathbf{y}_h - \mathbf{X}_h\boldsymbol{\beta})\},
\end{align*}
where $\mathbf{X}_h$ and $\mathbf{y}_h$ are the design matrix and response vector, respectively, of data provider $h$, $|.|$ is the matrix determinant and $\boldsymbol{\Sigma}_h = \boldsymbol{\Sigma}_h(\sigma^2,\mathbf{G}) = \mathbf{Z}_h\mathbf{G}\mathbf{Z}_h^T + \sigma^2I_{n_h}$. Due to the seemingly entangled data and parameter matrices, it is not straightforward to identify the aggregate statistics that can be used when the individual-level data are not available. Luo et al\citep{f6c8409ac3ed4b1cb27df3dfa95ac12c} showed that by utilizing the Woodbury matrix identity and some linear algebra concepts, the data can be disentangled from the parameters to reconstruct the profile log-likelihood. In their approach, only aggregate matrices $\mathbf{X}_h^T\mathbf{X}_h$, $\mathbf{X}_h^T\mathbf{y}_h$, $\mathbf{y}_h^T\mathbf{y}_h$, and $n_h$ from each data provider $h$ are required to produce identical estimates as those produced with the individual-level data. We have shown in the previous section that these aggregate data can actually be derived from the sample mean, sample covariance matrix, and sample size of each data set. Therefore, these summary statistics per data provider (e.g. per hospital) are also sufficient to estimate a linear mixed model in the absence of individual-level data. 

\subsection{Proposed method}

\noindent In this section, we provide details for creating the pseudo-data for a single variable and then extend it to a set of variables. As mentioned earlier, we opt to use these pseudo-data as input to the model estimation process rather than directly utilizing the sufficient statistics so that we can still use the existing linear regression or linear mixed model functionality in any statistical software such as the \texttt{lmer} function in the R package \texttt{lme4}.\cite{Rlme4} 

\subsubsection{Single variable}

The goal of constructing pseudo-data for linear models is to have exactly the same sample mean and variance as the original unavailable individual-level data. We do this by performing a linear transformation. For instance, suppose the original univariate sample $\mathbf{x}_d$ is unknown, but its sample mean $\bar{x}_d$ and sample standard deviation $s_d$ are available. We consider a linear transformation of a randomly generated data set $\mathbf{x}_r$ into pseudo-data $\mathbf{x}_\pi$ which has equal mean and standard deviation as $\mathbf{x}_d$. To this end, we let
\begin{align*}
    x_{\pi_i} &= a + bx_{r_i},
\end{align*}
where $\mathbf{x}_r$ can come from any distribution and has sample mean $\bar{x}_r$ and sample standard deviation $s_r$. Examining the relationship between the mean of the randomly generated data $\bar{x}_r$ and that of the transformed data $\bar{x}_\pi$, we find that
\begin{align*}
    \bar{x}_\pi = a + b\bar{x}_r,
\end{align*}
while for the variances:
\begin{align*}
    s_\pi^2 = b^2s_r^2,
\end{align*}
from which we find that to obtain identical means $\bar{x}_d = \bar{x}_\pi$ and standard deviations $s_d = s_\pi$ between the original unknown data $\mathbf{x}_d$ and the pseudo-data $\mathbf{x}_\pi$, we should let
\begin{align*}
    b &= \frac{s_\pi}{s_r} = \frac{s_d}{s_r}, \text{  and} \\
    a &= \bar{x}_\pi - \frac{s_\pi}{s_r}\bar{x}_r = \bar{x}_d - \frac{s_d}{s_r}\bar{x}_r, 
\end{align*}
so that
\begin{align*}
    x_{\pi_i} &= \bar{x}_d + s_d\frac{x_{r_i} - \bar{x}_r}{s_r}. 
\end{align*}
In summary, the algorithm to generate pseudo-data for a single variable is:
\begin{enumerate}
    \item Generate $n$ random numbers $\mathbf{x}_r$ from any distribution.
    \item Compute the sample mean $\bar{x}_r$ and sample standard deviation $s_{r}$ of $\mathbf{x}_r$.
    \item Transform $\mathbf{x}_r$ into the desired pseudo-data $\mathbf{x}_\pi$ using:
    \begin{align*}
    x_{\pi_i} &= \bar{x}_d + s_d\frac{x_{r_i} - \bar{x}_r}{s_r}, 
    \end{align*}
    where $\bar{x}_d$ and $s_d$ are the sample mean and sample standard deviation, respectively, of the original unavailable data.
\end{enumerate}

\subsubsection{Set of variables}
The strategy to construct pseudo-data for more than one variable is analogous to that for a single variable. Ripley \citep{Ripley87} presents an approach for multivariate normal distribution, but for our case, we do not have to impose normality on the pseudo-data nor on the initial set of random numbers. We just have to ensure that the resulting pseudo-data has exactly the same sample mean vector and sample covariance matrix as the individual-level data. Thus, from the algorithm for a single variable, we replace $\bar{x}_d$ with the sample mean vector for all variables ($\hat{\boldsymbol{\mu}}_d$). As for $s_d$, we take the Cholesky decomposition of the covariance matrix of all variables ($\hat{\boldsymbol{\Sigma}}_d$). Note that for a multiple linear regression or a linear mixed model, the set of variables comprises both the predictors and the response variable. The following algorithm provides a summary to generate pseudo-data with sample size $n$ for $p$ variables consisting of $p-1$ predictors and a continuous response variable:

\begin{enumerate}
    \item Generate random numbers $\mathbf{R} = [\mathbf{r}_1,...,\mathbf{r}_i,...\mathbf{r}_n]^T$ which is an $n\times p$ matrix where each column is generated independently from any distribution.
    \item Compute the mean vector $\hat{\boldsymbol{\mu}}_{r}$ and the covariance matrix $\hat{\boldsymbol{\Sigma}}_{r}$ of $\mathbf{R}$.
    \item Generate the $i$th pseudo-data point as
    $$\boldsymbol{\pi}_{i} = \hat{\boldsymbol{\mu}}_d + L_{\hat{\boldsymbol{\Sigma}}_d}(L_{\hat{\boldsymbol{\Sigma}}_{r}})^{-1}(\mathbf{r}_{i} - \hat{\boldsymbol{\mu}}_{r}),$$
    where $L_{\hat{\boldsymbol{\Sigma}}_d}$ and $L_{\hat{\boldsymbol{\Sigma}}_{r}}$ are the lower triangular matrices of the Cholesky decomposition of $\hat{\boldsymbol{\Sigma}}_d$ and $\hat{\boldsymbol{\Sigma}}_r$. Here, $\boldsymbol{\pi}_i$ is the pseudo-data vector consisting of the response variable and the predictors; that is, $[y_{\pi_i}, x_{\pi_{i1}}, ..., x_{\pi_{ij}}, ..., x_{\pi_{i(p-1)}}]^T$. 
\end{enumerate}

An issue that arises with using the Cholesky decomposition is the need to have a positive definite covariance matrix. In practical settings, especially when there are binary variables involved, this may not be true. An alternative procedure is to perform a singular value decomposition (SVD) on the centered values of $\mathbf{R}$ to obtain the matrix of right singular vectors denoted as $\mathbf{V}$. SVD can always be performed on any rectangular matrix such as this $n\times p$ matrix $\mathbf{R}$. The product of $\mathbf{R}$ and $\mathbf{V}$ is then computed and the resulting matrix elements are then divided by the root mean square $\sqrt{\sum_ix_{rv_{ij}}^2/(n-1)}$ of each column $j$. Denoting this as $\mathbf{R_V}$, the $i$th pseudo-data point is then generated as
\begin{align*}
    \boldsymbol{\pi}_i &= \hat{\boldsymbol{\mu}}_d + \mathbf{U}\boldsymbol{\Lambda}^{1/2}\mathbf{r_v}_i \hspace{1mm},
\end{align*}
where $\mathbf{U}$ is the matrix of eigenvectors of $\hat{\boldsymbol{\Sigma}}_d$ and $\boldsymbol{\Lambda}^{1/2}$ is a diagonal matrix whose elements are the square root of the eigenvalues of $\hat{\boldsymbol{\Sigma}}_d$. Eigendecomposition only requires that the matrix is diagonalized. Since a covariance matrix is always symmetric, it follows that it is always diagonalized and thus eigenvectors and eigenvalues can always be computed.\\ 

This alternative procedure is implemented by the function \texttt{mvrnorm} in the R package \texttt{MASS}\cite{RMASS} where each column of $\mathbf{R}$ is generated from a standard normal distribution. The user may wish to explore using a different distribution such as a uniform distribution to generate $\mathbf{R}$ by slightly altering the code for this function. Note that \texttt{mvrnorm} returns an error whenever the sample size of the pseudo-data $n$ is smaller than the number of variables $p$. The reason behind this is that the function \texttt{svd} used in \texttt{mvrnorm} returns a reduced matrix $\mathbf{V}$, affecting the dimension of the matrix $\mathbf{R_V}$ and making the matrix multiplication incompatible. We modified this setting so that the full SVD is returned. This R code is provided in Appendix \ref{modified_mvrnorm}.   

\section{An illustrative example: COVID-19 testing at CHOP}\label{section:example}
\noindent We demonstrate the proposed approach on a real dataset: the COVID-19 testing results from different clinics at the Children's Hospital of Pennsylvania (CHOP). It is a publicly available and deidentified dataset consisting of all patients who got tested at the hospital, and can be accessed from the R package \texttt{medicaldata} \citep{Rmedicaldata}. This dataset provides information about patients at CHOP who got tested for COVID-19 from days four to 107 of the pandemic in 2020. A total of 88 clinics from this hospital provided 15 524 patient records which were anonymized, time-shifted, and permuted. The COVID-19 test was performed via PCR. For a description of the variables, the reader may refer to the documentation of the \texttt{medicaldata} \citep{Rmedicaldata} package which is available online. Since we have access to the entire individual-level data, we will demonstrate how data providers can preprocess their raw data to achieve the expected summary data for our proposed method. In general, the data provider must supply the name and a brief description of each variable, the number of observations and mean per variable, and the covariance matrix. For model selection purposes, they must also provide the summary statistics for the standardized version, log- and squared transformations of the numeric variables, as well as for the two-way interactions among the variables. We illustrate these using R software, but these results can also be implemented using any other statistical software.\\  

Suppose we are interested in how factors such as gender, age, and whether the specimen was collected in a drive-thru site affect the cycle at which threshold reached during PCR (numeric variable from 14 to 45). For this data set, the cycle threshold is the response variable while gender, age, drive thru indicator, and an interaction term for age and gender are the predictors. The cycle at which threshold reached during PCR is a measure of how much amplification is necessary to detect the target viral gene, and is inversely proportional to viral load.\citep{Rao2020} This means that if more cycles are needed to detect a viral gene, then the presence of the viral gene is less likely. Some studies found a negative correlation between the cycle threshold and disease severity \citep{10.1093/ofid/ofab453} and mortality among patients.\citep{10.1371/journal.pone.0244777} In this illustrative example, we will model how the aforementioned regressors influence this measure to hypothesize about the clinical outcomes among patients at CHOP.

\subsection{Preliminary analysis by the data provider}
Each data provider should supply the data analyst with a good overview of all the available variables. The data provider may use the function \texttt{skim} from the R package \texttt{skimr}\cite{Rskimr} to accomplish this. Since the CHOP data from R is already composed of the pooled individual-level records from all 88 clinics, applying the aforementioned function covers all clinics already. Table \ref{tab:chop_skim} displays the metadata for all 88 clinics having a total of 15 524 patient records during the COVID-19 pandemic in 2020.    \\

\begin{table}[h!]
    \centering
    \caption{Metadata of all clinics at CHOP}
    \label{tab:chop_skim}
    \resizebox{\textwidth}{!}{
    \begin{tabular}{lrrrrrrrr}
    \hline
        Variable & Type & Number of & Complete & Number of& Number of  \\
        name& &missing observations & rate & empty cells & unique values \\
    \hline
        Fake first name & char & 0 & 1.00 & 0 & 832 \\
        Fake last name & char & 0 & 1.00 & 0 & 27 \\
        Gender & char & 0 & 1.00 & 0 & 2 \\
        Test ID & char & 0 & 1.00 & 0 & 2 \\
        Clinic name & char & 0 & 1.00 & 0 & 88 \\
        Result & char & 0 & 1.00 & 0 & 3 \\
        Demographic group & char & 0 & 1.00 & 0 & 5 \\
        Payor group & char & 7087 & 0.54 & 0 & 7 \\
        Patient class & char & 7077 & 0.54 & 0 & 9 \\
        Subject ID & num & 0 & 1.00 & - & - \\
        Day of & & & & & \\
        pandemic & num & 0 & 1.00 & - & - \\
        Age & num & 0 & 1.00 & - & - \\
        Drive thru &  & &  &  &  \\
        indicator & num & 0 & 1.00 & - & - \\
        Cycle threshold &  & &  &  &  \\
        result & num & 209 & 0.99 & - & - \\
        Orderset & num & 0 & 1.00 & - & -\\
        Collection to & & & & & & \\
        receive time & num & 0 & 1.00 & - & - \\
        Receive to & & & & & & \\
        verification time & num & 0 & 1.00 & - & - \\
    \hline
    \end{tabular}}
\end{table}

\normalsize
Our proposed method assumes that the summary data per data provider were computed from complete observations. Additionally, for categorical variables, the data provider must also indicate the levels. For instance, for binary variables such as gender, the variable name in the summary data must reflect the non-reference category (e.g. gendermale). For this dataset, of the 88 clinics, only 70 clinics with a total of 15 068 observations were included in the analysis after filtering out incomplete observations and invalid values such as NA. Those clinics with only one observation were removed as well because in a federated data setting, it is not so common for a data provider (e.g. a clinic) to have only one patient. Moreover, even if a clinic with only one patient exists, summarizing the data does not make sense and neither does handing over the single patient record to the data analyst because it goes against that patient's right to data confidentiality. Should the clinic do so after fulfilling some legal requirements to ensure privacy, this observation itself can be combined with the generated pseudo-data. In Appendix \ref{chop_app}, we provide an R code for preprocessing this data set. As an example, Table \ref{tab:chop_summary_stat} displays the summary statistics for the Inpatient Ward A clinic at CHOP as well as the covariance matrix for the variables we are including in our model. \\

\begin{table}[h!]
\caption{Summary statistics for the Inpatient Ward A clinic at CHOP}
\resizebox{\textwidth}{!}{
\begin{tabular}{lrrrrrrr}
\hline
Variable & n & Mean & \multicolumn{5}{c}{Variance-Covariance}\\
name & & & log of Cycle & Gendermale & Age & Drive thru & Gendermale $\times$ \\
& & & threshold & & &  & Age \\
\hline
log of Cycle threshold & 208 & 3.803 & 0.001 & 0.001 & 0.001 & 0.000 & -0.001\\
Gendermale & 208 & 0.529 & 0.001 & 0.250 & 0.053 & 0.002 & 0.369\\
Age & 208 & 1.373 & 0.001 & 0.053 & 10.506 & 0.003 & 6.621\\
Drive thru & 208 & 0.005 & 0.000 & 0.002 & 0.003 & 0.005 & 0.006\\
Gendermale $\times$ Age & 208 & 0.779 & -0.001 & 0.369 & 6.621 & 0.006 & 7.085\\
\hline
\end{tabular}}
\label{tab:chop_summary_stat}
\end{table}

\subsection{LMM estimation by the data analyst}
After receiving the sufficient statistics from the data providers, the data analyst formulates the linear mixed model. To study the relationship between COVID-19 PCR test cycle threshold and patient information namely gender, age, their interaction, and the drive thru indicator, we fit two linear mixed models: one with only a random intercept per clinic and another with a random intercept and a random slope for age. This allows variations in mean cycle threshold and age effects across the clinics. We use the logarithm of the cycle threshold to ensure nonnegative values for this variable and we standardize age to avoid numerical problems during the optimization of the log-likelihood. One set of pseudo-data was generated for each clinic using the proposed method. Using the \texttt{lmer} function in the R package \texttt{lme4} on the pseudo-data and on the actual data, we estimate the parameters of the model. Table \ref{tab:chop_lmm_randomint} displays the results for the model with only a random intercept while Table \ref{tab:chop_lmm_randomslope} presents the estimates for an LMM with an additional random slope for age. For both models, we find that only the scaled residuals are different between the LMM using pseudo-data and using actual data. This is expected since residuals are computed from individual observations and thus cannot be reproduced from a different set of values such as the pseudo-data. Additionally, we can reproduce the AIC and confidence intervals, as shown in the same tables. We select the model with both random intercept and random slope since it has lower AIC and BIC values.\\

From the estimates of the selected model, we observe that among the fixed effects, the interaction of gender and age significantly affect the cycle threshold, whereas drive-thru testing does not. This significant interaction effect indicate that the overall impact of a patient's gender on the log cycle threshold also depends on the age group of the patient, and vice versa. Moreover, since the effects of age are allowed to vary across clinics, the interpretation also varies per clinic. For instance, for a male patient who belongs to clinic $h$ and is one standard deviation or around 16 years older, the log cycle threshold changes by $-0.0057 + b_{age_h}$. If he belongs to the Inpatient Ward A, the log cycle threshold decreases by $0.008$. On the other hand, a female patient belonging to the same clinic is expected to have $0.003$ decrease in log cycle threshold for every one standard deviation increase in age. This suggests that for this clinic, older patients tend to have more viral load, although the difference across age groups is slightly more pronounced among males and among females. Note that this per clinic interpretation is also supported by our proposed approach since the random effects prediction are exactly equal to those derived from the actual data.. Lastly, the estimated variance components of the random effects suggest that there is not much variation across the clinics. \\

\begin{table}
\caption{Comparison of LMMs with random intercept only based on pseudo-data and based on actual CHOP data.}
\begin{tabular}{lrr}
\hline
 & \multicolumn{2}{c}{LMM estimates}\\
 & \multicolumn{1}{c}{pseudo-data} & \multicolumn{1}{c}{actual data} \\
 \hline
 REML criterion at convergence & -20473 & -20473\\
 Scaled residuals: & & \\
 \multicolumn{1}{c}{Min} & -4.5919 & -9.3287\\
 \multicolumn{1}{c}{$Q_1$} & -0.5773 & 0.1482\\
 \multicolumn{1}{c}{Median} & 0.0249 & 0.2174\\
 \multicolumn{1}{c}{$Q_3$} & 0.5760 & 0.2538\\
 \multicolumn{1}{c}{Max} & 3.7464  & 1.1855\\
\hline
(Intercept)                   & 3.7871 (0.0039){***} & 3.7871 (0.0039){***} \\
Gendermale                    & 0.0021 (0.002){\textcolor{white}{***}}  & 0.0021 (0.002){\textcolor{white}{***}}\\
Std. Age                           & -0.0046 (0.0015){\textcolor{white}{*}**} &-0.0046 (0.0015){\textcolor{white}{*}**} \\
Drive thru      & -0.0043 (0.0058){\textcolor{white}{***}} & -0.0043 (0.0058){\textcolor{white}{***}}\\
Gendermale $\times$ Std. Age                & -0.0061 (0.0020){\textcolor{white}{*}**} & -0.0061 (0.0020){\textcolor{white}{*}**}\\
\hline
AIC                           & -20459.04              & -20459.04              \\
BIC                           & -20405.7              & -20405.7              \\
n                     & 15068                    & 15068                    \\
number of clinics     & 70                       & 70                       \\
$\sigma_{Int}$ & 0.0216                   & 0.0216                   \\
$\sigma$                 & 0.1222                   & 0.1222                  \\
\hline
\end{tabular}
\begin{tabular}{lrrrr}
& \multicolumn{4}{c}{95\% confidence bounds}\\
& \multicolumn{2}{c}{pseudo-data} & \multicolumn{2}{c}{actual data}\\
  & 2.5 \% & 97.5 \% & 2.5 \% & 97.5 \%\\
\hline
$\sigma_{Int}$ & 0.0160 & 0.0282 & 0.0160 & 0.0282\\
$\sigma$ & 0.1208 & 0.1235 & 0.1208 & 0.1235\\
(Intercept) & 3.7793 & 3.7949 & 3.7793 & 3.7949\\
Gendermale & -0.0018 & 0.006 & -0.0018 & 0.006\\
Std. Age & -0.0076 & -0.0015 & -0.0076 & -0.0015\\
Drive thru & -0.0156 & 0.0071 & -0.0156 & 0.0071\\
Gendermale $\times$ Std. Age & -0.0100 & -0.0022 & -0.0100 & -0.0022\\
\hline
\multicolumn{3}{l}{\scriptsize{$^{***}p<0.001$; $^{**}p<0.01$; $^{*}p<0.05$}}
\end{tabular}
\label{tab:chop_lmm_randomint}
\end{table}

\begin{table}
\caption{Comparison of LMMs with random intercept and random slope for age based on pseudo-data and based on actual CHOP data.}
\begin{tabular}{lrr}
\hline
 & \multicolumn{2}{c}{LMM estimates}\\
 & \multicolumn{1}{c}{pseudo-data} & \multicolumn{1}{c}{actual data} \\
 \hline
 REML criterion at convergence & -20513.2 & -20513.2\\
 Scaled residuals: & & \\
 \multicolumn{1}{c}{Min} & -4.7364 & -9.3604\\
 \multicolumn{1}{c}{$Q_1$} & -0.5789 & 0.1229\\
 \multicolumn{1}{c}{Median} & 0.0241 & 0.2030\\
 \multicolumn{1}{c}{$Q_3$} & 0.5777 & 0.2560\\
 \multicolumn{1}{c}{Max} & 3.7682 & 1.8698\\
\hline
(Intercept)                   & 3.7851 (0.0045){***} & 3.7851 (0.0045){***} \\
Gendermale                    & 0.0021 (0.0020){\textcolor{white}{***}}  & 0.0021 (0.0020){\textcolor{white}{***}}\\
Std. Age                           & -0.0005 (0.0037){\textcolor{white}{***}} &-0.0005 (0.0037){\textcolor{white}{***}}\\
Drive thru      & -0.0038 (0.0059){\textcolor{white}{***}}      & -0.0038 (0.0059){\textcolor{white}{***}}\\
Gendermale $\times$ Std. Age                & -0.0052 (0.0020){\textcolor{white}{*}**} & -0.0052 (0.0020){\textcolor{white}{*}**}\\
\hline
AIC                           & -20495.15              & -20495.15              \\
BIC                           & -20426.57              & -20426.57              \\
n                     & 15068                    & 15068                    \\
number of clinics     & 70                       & 70                       \\
$\sigma_{Int}$ & 0.0249                   & 0.0249                   \\
$\sigma_{Age}$ & 0.0128                   & 0.0128                   \\
$\sigma_{\text{Int} \times \text{Age}}$ & -0.10 & -0.10 \\
$\sigma$                 & 0.1219                   & 0.1219                   \\
\hline
\end{tabular}
\begin{tabular}{lrrrr}
& \multicolumn{4}{c}{95\% confidence bounds}\\
& \multicolumn{2}{c}{pseudo-data} & \multicolumn{2}{c}{actual data}\\
  & 2.5 \% & 97.5 \% & 2.5 \% & 97.5 \%\\
\hline
$\sigma_{Int}$ & 0.0185 & 0.0323 & 0.0185 & 0.0323\\
$\sigma_{\text{Int} \times \text{Age}}$ & -0.5797 & 0.3917 & -0.5797 & 0.3917\\
$\sigma_{Age}$ & 0.008 & 0.019 & 0.008 & 0.019\\
$\sigma$ & 0.1205 & 0.1233 & 0.1205 & 0.1233\\
(Intercept) & 3.7762 & 3.7942 & 3.7762 & 3.7942\\
Gendermale & -0.0018 & 0.006 & -0.0018 & 0.006\\
Age & -0.0081 & 0.0074 & -0.0081 & 0.0074\\
Drive thru & -0.0152 & 0.0077 & -0.0152 & 0.0077\\
Gendermale $\times$ Age & -0.0092 & -0.0013 & -0.0092 & -0.0013\\
\hline
\multicolumn{3}{l}{\scriptsize{$^{***}p<0.001$; $^{**}p<0.01$; $^{*}p<0.05$}}
\end{tabular}
\label{tab:chop_lmm_randomslope}
\end{table}


\section{Discussion}\label{section:discussion}
\noindent In this paper, we have demonstrated that data privacy in a federated data setting can be achieved not only through machine learning algorithms but even through age-old statistical concepts such as sufficiency and likelihood. These principles enable data reduction without losing important information about the parameter of interest. Specifically, since the sufficient statistics already contain the information required to estimate the parameters of a statistical model, data providers do not have to share individual observations anymore. This approach is very useful and applicable to settings where individual-level data are too sensitive to be shared, such as patient data from hospitals. For a linear regression model and a linear mixed model, this is true since their sufficient statistics exist. For other more complex models such as generalized linear models with and without random effects, this may not be as straightforward.  \\

Since our approach produces identical estimates as those obtained from pooling the individual-level records across multiple data providers, we have confidence that our estimates are as good as the estimates that the established estimation techniques claim. Among the methods proposed in the literature, maximum likelihood (ML) and residual or restricted maximum likelihood (REML) have become the standard methods for estimating the parameters of a linear mixed model.\citep{GUMEDZE20111920} However, between these two, ML estimators do not account for the degrees of freedom lost when estimating the fixed effects, resulting in biased variance parameter estimates towards the null, especially for small samples.\citep{LIN19842389, Swallow_Monahan_1984} For this reason, REML estimation of the variance parameters is preferred over ML. Moreover, REML estimators have shown improved properties whenever the number of clusters is small,\citep{Ferron2009, doi:10.1080/00273171.2016.1167008} which suffers from finite sample bias more than models involving small cluster sizes.\citep{2bc55cd7-0921-340f-ac02-0e3129338343} Despite these desirable properties, REML does not completely solve the issues related to inflated Type I error rates for fixed effects.\citep{doi:10.1080/00273171.2017.1344538} To address this, the Kenward-Roger correction has been recommended as best practice in the literature since it has been shown to maintain nominal Type I error rates.\citep{Ferron2009} Due to the nature of our proposed strategy, versatility in specifying the estimation procedure (ML or REML) and applying corrections (e.g. Kenward-Roger) whenever the sample size is small can be easily implemented with identical results as with the actual observations. \\

In contrast to the study of Luo et al \cite{f6c8409ac3ed4b1cb27df3dfa95ac12c} which also yields exactly the same estimates for LMM, our approach is a simple one and thus can more easily be implemented in practice. Another advantage of our proposed framework is that we do not need to specify a distribution from which the pseudo-data come from, unlike the method proposed by Song et al.\cite{Song2020-um} Additionally, the concept can be applied using any statistical software that can estimate LMM, thus enabling a wider scope of implementation. Another edge we have is the computational efficiency of generating only one set of pseudo-data compared to methodologies that simulate data multiple times and aggregate the estimates to form a single parameter estimate. As a consequence, we are spared from the question of how many simulations to run and which aggregation method to best implement. Lastly, in contrast to federated learning algorithms in the literature, our approach does not require more than one communication iteration between the data providers and  data analyst, nor do we need to set up a network among the databases. Hence, we are significantly minimizing, if not totally eliminating, the risk of disclosing sensitive data.   \\

A limitation of our proposed approach is the inability to compute residuals, which require individual response values from the original data. The pseudo-data we generate, although similar in some characteristics to the original, cannot be used to compute residuals. Thus, model diagnostics through residual plots cannot be performed. In general, even when the individual-level data are available, model checking through residual analysis is a non-trivial task.\citep{Bradburn2003, lindsey2001nonlinear} This is especially true for visual assessment because of the element of subjectivity and the difficulty in discerning patterns when the sample size is very large, which is important when deciding possible corrective measures.\citep{harrell2015regression} Formal tests on residuals, on the other hand, can become very sensitive to large samples, which may lead to falsely concluding violations of the assumption. Hence, a good recourse is to be aware of (1) the consequences of potential violations of the model assumptions and (2) possible remedies to mitigate these consequences. \\

To begin with, the normality assumption regarding the response variable has the least impact on tests and inference derived from linear regression \citep{ALI1996175, ba2e7a12-bb49-3394-a495-cce958f9588f, Gelman_Hill_2006} and linear mixed models \citep{https://doi.org/10.1111/2041-210X.13434} as long as outliers are handled properly. Specifically for linear mixed models, inference on the fixed effects remain valid even when the random effects do not follow a normal distribution.\citep{gałecki2013linear} Gelman and Hill\cite{Gelman_Hill_2006} do not recommend checking the residuals for normality while Galecki and Burzykowski \cite{gałecki2013linear} note that normality is not important for ordinary least squares although it is for maximum likelihood estimation. On the contrary, heteroscedasticity affects the standard errors of the parameter estimates of a classical linear regression model even though the point estimates remain unbiased.\citep{380055f1-819e-3149-a01c-cfab4e53c79f} This affects confidence interval construction as well as inference about the covariate effects. One way to address this without altering the interpretation of effects through variable transformation \citep{weisberg2005applied, carroll1988transformation} is by using robust standard errors.\citep{f457a0f7-5c1e-3d02-9161-7806a34faad9} We can show that with additional summary statistics namely the third and fourth joint sample moments, our approach can also achieve identical robust variance estimates as the ones derived from actual data (Appendix \ref{robvar}). Analogously, a robust or empirical variance estimator has been proposed for linear mixed models and has been shown to be consistent under misspecification of the correlation structure as long as the mean is correctly specified.\citep{10.1093/biomet/73.1.13} \\

Validity, additivity and linearity, and independence of errors are the top most important assumptions when utilizing linear regression models.\citep{Gelman_Hill_2006} Although these cannot be evaluated in our proposed framework, model selection procedures may help mitigate the impact of violations to these assumptions, if they exist. Harrell\cite{harrell2015regression} proposed fitting a flexible parametric model that allows for most departures from the assumptions as an alternative to residual analysis. In light of our proposed strategy, selecting from candidate models that consider potential violations is recommended in practice (e.g. considering different combination of regressors, polynomial terms). The independence assumption may be relaxed by considering a random effects model instead of the classical linear regression model.\\

Another consequence of our approach is the inability to perform training and testing since partitioning the original data is not possible. As a result, model validation would not be an option, and predictive accuracy of the resulting model might be difficult to assess. A potential remedy is to generate pseudo-data which embodies the statistical properties of the original data more closely than just having identical summary statistics such as the mean vector and covariance matrix. Several studies present different strategies to generate and analyze synthetic data from a statistical disclosure control perspective as well as from a machine learning perspective \citep{doi:10.1146/annurev-statistics-040720-031848, HERNANDEZ202228, math10152733, MURTAZA2023100546}, but these would of course not be as simple anymore and would require more data processing from the data providers' end. \\

An interesting point to consider is the impact of rounding off sufficient statistics. In the current implementation of the proposed framework, we assume that the mean vector and covariance matrix from each data provider contain the exact values and not the rounded off values. In practice, these sufficient statistics may be rounded off to a few decimal places. To examine this, we implemented the proposed approach on sufficient statistics that were rounded off to two decimal places. We observed that there were only slight differences in the model estimates when compared to using the exact values of the sufficient statistics (e.g. the difference starts from the third decimal place). The direction of effect and the overall inference also remained the same. This suggests that the method is robust to rounded values of the sufficient statistics in this setting, although a more thorough sensitivity analysis is encouraged to draw more conclusive findings. \\

A field related to federated learning is meta-analysis. Like federated learning, meta-analysis aims to build a global model that synthesizes information from multiple studies. Since meta-analysis dates back to as early as 1976 \citep{doi:10.3102/0013189X005010003} while federated learning is fairly recent,\citep{45648} it is worthwhile to explore meta-analysis techniques in addressing the challenges of a federated data setting. Traditionally, meta-analysis directly utilizes the aggregated information from studies. These techniques are called aggregate data meta-analysis (ADMA). However, individual participant data meta-analysis (IPDMA) is now regarded as the ``gold standard", but accessing individual-level data remains an obstacle.\citep{Nevittj1390} The work of Papadimitropoulou et al\cite{10.1002/jrsm.1331} involving pseudo-IPD thus provides a good solution to performing IPDMA even when studies only include aggregate information. Because of its similarity to a meta-analysis setting, the federated data setting also benefits from this framework. In contrast to a meta-analysis setting though, multiple variables can be included in a federated data setting because data providers may be more willing to share the covariance structure of the variables.  \\

Some future research avenues include dealing with missing data and generating sufficient statistics for interaction terms and transformed variables e.g. log- and quadratic-transformed variables, from the sufficient statistics of the main variables. Presently, our approach assumes that the data providers hand over summary data for complete observations. They must also supply the sufficient statistics for the transformed variables and for possible interaction terms on top of those for the main variables. Including random slopes and more complex hierarchical models is also a viable direction. Currently, the authors are working on generalized linear models with and without random intercept wherein we find the potential of extending the idea of pseudo-data generation that matches the summary statistics of the actual data.

\section{Conclusion}
\noindent In this paper, we have demonstrated that parameter estimation of a linear mixed model can be performed on federated data by generating pseudo-data from the sample size, mean vector, and covariance matrix supplied by each data provider. The principles of statistical sufficiency and likelihood provide a good theoretical support to the validity of the proposed framework. Estimates achieved from this approach are identical to those obtained from the actual individual-level data, which are difficult to access due to privacy reasons. Simplicity, computational and communication efficiency, and potentially wider scope of implementation through any statistical software distinguish our approach from the existing strategies in the literature. Extending this approach to generalized linear mixed models is a current work in progress.

\section{Conflict of Interest}
The author(s) declare(s) no potential conflicts of interest with respect to the research, authorship and/or publication of this article.


\section{Funding}
    The author(s) disclosed receipt of the following financial support for the research, authorship, and/or publication of this article: The author(s) acknowledge(s) support from the Methusalem financement program of the Flemish Government.



\appendix
\section{Appendix}
\subsection{Likelihood principle}\label{lik_prin}
Another principle of data reduction discussed by \cite{CaseBerg:01} is the likelihood principle. Stated more formally, 
\begin{quote}
    ``If $\mathbf{x_1}$ and $\mathbf{x_2}$ are two samples such that the likelihood $L(\theta|\mathbf{x_1})$ is proportional to $L(\theta|\mathbf{x_2})$, that is, there exists a constant $C(\mathbf{x_1}, \mathbf{x_2})$ such that
$$L(\theta|\mathbf{x_1}) = C(\mathbf{x_1}, \mathbf{x_2})L(\theta|\mathbf{x_2})$$
then the conclusions drawn from $\mathbf{x_1}$ and $\mathbf{x_2}$ should be identical.''
\end{quote}

They demonstrated this principle for the case of a normal distribution and showed that the likelihood of a parameter $\mu$ given a sample $\mathbf{x_1}$ ($L(\mu|\mathbf{x_1})$) can be exactly equal to the likelihood of the same parameter given another sample $\mathbf{x_2}$ ($L(\mu|\mathbf{x_2})$) if their sample means are equal ($\mathbf{\bar{x}}_1 = \mathbf{\bar{x}}_2$). 

\subsection{Showing the sufficient statistics for a linear regression model}\label{suf_lin_reg}
Given $n$ observations, an intercept, and $p-1$ predictors, if $\mathbf{X}$ denotes the $n \times p$ design matrix and $\mathbf{y}$ is the $n \times 1$ vector of continuous responses, the linear regression coefficients $\boldsymbol{\beta}$ and the variance $\sigma^2$ can be estimated through the log-likelihood
\begin{align*}
    l(\boldsymbol{\beta}, \sigma^2; \mathbf{y}, \mathbf{X}) &= -\frac{n}{2}\text{ln}(2\pi) -\frac{n}{2}\text{ln}(\sigma^2) -\frac{1}{2\sigma^2}\sum_{i=1}^n(y_i-\mathbf{x}_i^T\boldsymbol{\beta})^2
\end{align*}
where $\mathbf{x}_i$ is a vector containing the $i$th row in the design matrix. We see here that information from the sample is required only in the last term. Moreover, the sum of squares of this term can be expressed as
\begin{align*}
    \sum_{i=1}^n(y_i-\mathbf{x}_i^T\boldsymbol{\beta})^2 &= \sum_{i=1}^n(y_i^2 - 2y_i\mathbf{x}_i^T\boldsymbol{\beta} + (\mathbf{x}_i^T\boldsymbol{\beta})(\mathbf{x}_i^T\boldsymbol{\beta})^T). 
\end{align*}
Since $\mathbf{x}_i^T\boldsymbol{\beta}$ is just the dot product of two vectors $\mathbf{x}_i$ and $\boldsymbol{\beta}$, commutativity applies such that the equation above can also be written as
\begin{align*}
    \sum_{i=1}^n(y_i-\mathbf{x}_i^T\boldsymbol{\beta})^2 &= \sum_{i=1}^n(y_i^2 - 2y_i\mathbf{x}_i^T\boldsymbol{\beta} + (\boldsymbol{\beta}^T\mathbf{x}_i)(\boldsymbol{\beta}^T\mathbf{x}_i)^T) 
\end{align*}
which when simplified further yields
\begin{align*}
    \sum_{i=1}^n(y_i-\mathbf{x}_i^T\boldsymbol{\beta})^2 &= \sum_{i=1}^n(y_i^2 - 2y_i\mathbf{x}_i^T\boldsymbol{\beta} + \boldsymbol{\beta}^T\mathbf{x}_i\mathbf{x}_i^T\boldsymbol{\beta}) \\
    &= \sum_{i=1}^ny_i^2 - 2\sum_{i=1}^ny_i\mathbf{x}_i^T\boldsymbol{\beta} + \sum_{i=1}^n\boldsymbol{\beta}^T\mathbf{x}_i\mathbf{x}_i^T\boldsymbol{\beta}
\end{align*}

From this, we find that knowing $n$, $\sum_{i=1}^ny_i^2$, $\sum_{i=1}^ny_i\mathbf{x}_i^T$, and $\sum_{i=1}^n\mathbf{x}_i\mathbf{x}_i^T$ is sufficient to construct the log-likelihood and estimate the parameters, even in the absence of individual-level data. In particular, $\sum_{i=1}^ny_i\mathbf{x}_i^T$ and $\sum_{i=1}^n\mathbf{x}_i\mathbf{x}_i^T$ are sufficient to estimate the coefficients $\boldsymbol{\beta}$ while the variance $\sigma^2$ also requires $\sum_{i=1}^ny_i^2$ in addition to the other two. Furthermore, these values can be obtained from the vector of sample means and sample covariance matrix of the response variable and the predictors. Specifically, since the sample variance $s_{\mathbf{y}}^2$ is computed as
\begin{align*}
    s^2_{\mathbf{y}} &= \frac{1}{n-1}\sum_{i=1}^n(y_i - \bar{\mathbf{y}})^2 \\
          &= \frac{1}{n-1}\sum_{i=1}^n(y_i^2 - 2y_i\bar{\mathbf{y}} + \bar{\mathbf{y}}^2)\\
          &= \frac{1}{n-1}\left(\sum_{i=1}^ny_i^2 - 2\bar{\mathbf{y}}\sum_{i=1}^ny_i + n\bar{\mathbf{y}}^2\right)\\
          &= \frac{1}{n-1}\left(\sum_{i=1}^ny_i^2 - n\bar{\mathbf{y}}^2\right),
\end{align*}
performing some algebraic manipulations will show that $\sum_{i=1}^ny_i^2$ can be derived from the sample variance $s^2_{\mathbf{y}}$, the sample mean $\bar{\mathbf{y}}$, and the sample size $n$
\begin{align*}
    \sum_{i=1}^ny_i^2 &= s^2_{\mathbf{y}}(n-1) + n\bar{\mathbf{y}}^2.
\end{align*}
For $\sum_{i=1}^ny_i\mathbf{x}_i^T$, we note that $y_i\mathbf{x}_i^T$ is a $1\times p$ matrix
\begin{align*}
    \begin{bmatrix}
        y_i & y_ix_{i1} & y_ix_{i2} & ... & y_ix_{ij} & ... & y_ix_{i(p-1)} 
    \end{bmatrix}
\end{align*}
where the first element of $\mathbf{x}_i$ is 1 corresponding to the intercept. Thus, $\sum_{i=1}^ny_i\mathbf{x}_i^T$ is a $1\times p$ matrix
\begin{align*}
    \begin{bmatrix}
        \sum_{i=1}^ny_i & \sum_{i=1}^ny_ix_{i1} & \sum_{i=1}^ny_ix_{i2} & ... & \sum_{i=1}^ny_ix_{ij} & ... & \sum_{i=1}^ny_ix_{i(p-1)} 
    \end{bmatrix}.
\end{align*}
The first element can be obtained from the sample mean $\bar{\mathbf{y}}$ while the rest of the elements needs the sample covariance between $\mathbf{y}$ and each of the predictors:
\begin{align*}
    s_{\mathbf{yx}_j} &= \frac{1}{n-1}\sum_{i=1}^n(y_i-\bar{\mathbf{y}})(x_{ij}-\bar{\mathbf{x}}_j) \\
    &= \frac{1}{n-1}\sum_{i=1}^n(y_ix_{ij}-\bar{\mathbf{y}}x_{ij}-y_i\bar{\mathbf{x}}_j+\bar{\mathbf{y}}\bar{\mathbf{x}}_j)\\
    &= \frac{1}{n-1}\left(\sum_{i=1}^ny_ix_{ij}-\bar{\mathbf{y}}\sum_{i=1}^nx_{ij}-\bar{\mathbf{x}}_j\sum_{i=1}^ny_i+n\bar{\mathbf{y}}\bar{\mathbf{x}}_j\right)\\
    &= \frac{1}{n-1}\left(\sum_{i=1}^ny_ix_{ij}-\bar{\mathbf{y}}\sum_{i=1}^nx_{ij}\right),
\end{align*}
and thus,
\begin{align*}
    \sum_{i=1}^n{y_ix_{ij}} &= s_{\mathbf{yx}_j}(n-1) + \bar{\mathbf{y}}\sum_{i=1}^nx_{ij}.
\end{align*}
Lastly, for $\sum_{i=1}^n\mathbf{x}_i\mathbf{x}_i^T$, each summand $\mathbf{x}_i\mathbf{x}_i^T$ yields a $p\times p$ matrix
\begin{align*}
    \begin{bmatrix}
        1 & x_{i1} & \hdots & x_{i(p-1)} \\
        x_{i1} & x_{i1}^2 & \hdots & x_{i1}x_{i(p-1)}\\
        \vdots & \vdots & \ddots & \vdots\\
        x_{i(p-1)} & x_{i(p-1)}x_{i1} & \hdots & x_{i(p-1)}^2 
    \end{bmatrix}
\end{align*}
which when summated over $n$ observations yields
\begin{align*}
    \begin{bmatrix}
        n & \sum_i{x_{i1}} & \hdots & \sum_i{x_{i(p-1)}} \\
        \sum_i{x_{i1}} & \sum_i{x_{i1}^2} & \hdots & \sum_i{x_{i1}x_{i(p-1)}}\\
        \vdots & \vdots & \ddots & \vdots\\
        \sum_i{x_{i(p-1)}} & \sum_i{x_{i(p-1)}x_{i1}} & \hdots & \sum_i{x_{i(p-1)}^2} 
    \end{bmatrix}.
\end{align*}
Performing similar derivations as above reveals that computing $\sum_{i=1}^n\mathbf{x}_i\mathbf{x}_i^T$ only requires the sample mean ($\bar{\mathbf{x}}_j$), variance ($s^2_{\mathbf{x}_j}$), and covariances among predictors $j$ and $k$ ($s_{\mathbf{x}_j\mathbf{x}_k}$), namely
\begin{align*}
    \sum_{i=1}^nx_{ij}^2 &= s^2_{\mathbf{x}_j}(n-1) + n\bar{\mathbf{x}}_j^2, \\
    \sum_{i=1}^n{x_{ij}x_{ik}} &= s_{\mathbf{x}_j\mathbf{x}_k}(n-1) + \bar{\mathbf{x}}_j\sum_{i=1}^nx_{ik}.
\end{align*}

\subsection{Robust variance estimation from summary statistics}\label{robvar}
When estimating the robust variance of linear regression coefficients, the following estimator is used when the individual observations are available:

        \begin{align*}
            \hat{V}(\hat{\boldsymbol{\beta}}) &= (\mathbf{X}^T\mathbf{X})^{-1}
            (\mathbf{X}^T\mathbf{W}\mathbf{X})^{-1}(\mathbf{X}^T\mathbf{X})^{-1} 
        \end{align*}

        where $\mathbf{X}$ is the design matrix and $\mathbf{W}$ is an $n\times n$ diagonal matrix whose elements consist of the squared residuals $\hat{e}_i^2 = (y_i-\mathbf{x}_i^T\boldsymbol{\beta})^2$. $\mathbf{X}^T\mathbf{X}$ is a $p \times p$ matrix that is equivalent to $\sum_{i=1}^{n}\mathbf{x}_i\mathbf{x}_i^T$, which we have shown (Appendix \ref{suf_lin_reg}) can be computed from the mean vector and covariance matrix of the variables. Recall that $\mathbf{x}_i, i=1,..,n$ denotes the vector representing the $i$th row of $\mathbf{X}$. On the other hand, $(\mathbf{X}^T\mathbf{W}\mathbf{X})^{-1}$ can be shown to be composed of summary statistics involving the third and fourth joint sample moments. Specifically,
        \begin{align*}
            (\mathbf{X}^T\mathbf{W}\mathbf{X})^{-1} =& \sum_{i=1}^n\hat{e}_i^2\mathbf{x}_i\mathbf{x}_i^T\\
            =& \sum_{i=1}^n(y_i-\mathbf{x}_i^T\boldsymbol{\beta})^2\mathbf{x}_i\mathbf{x}_i^T\\
            =& \sum_{i=1}^n(y_i^2\mathbf{x}_i\mathbf{x}_i^T -2(y_i\mathbf{x}_i^T\boldsymbol{\beta})\mathbf{x}_i\mathbf{x}_i^T + (\boldsymbol{\beta}^T\mathbf{x}_i\mathbf{x}_i^T\boldsymbol{\beta})\mathbf{x}_i\mathbf{x}_i^T)
        \end{align*}

        The first term $y_i^2\mathbf{x}_i\mathbf{x}_i^T$ is just a product of a scalar $y_i^2$ and the matrix $\mathbf{x}_i\mathbf{x}_i^T$, which results in the matrix
        \begin{align*}
            \begin{bmatrix}
        y_i^2 & y_i^2x_{i1} & \hdots & y_i^2x_{i(p-1)} \\
        y_i^2x_{i1} & y_i^2x_{i1}^2 & \hdots & y_i^2x_{i1}x_{i(p-1)}\\
        \vdots & \vdots & \ddots & \vdots\\
        y_i^2x_{i(p-1)} & y_i^2x_{i(p-1)}x_{i1} & \hdots & y_i^2x_{i(p-1)}^2 
            \end{bmatrix}
        \end{align*} 
        whose summation over all observations $i$ becomes
        \begin{align*}
            \begin{bmatrix}
        \sum_iy_i^2 & \sum_iy_i^2x_{i1} & \hdots & \sum_iy_i^2x_{i(p-1)} \\
        \sum_iy_i^2x_{i1} & \sum_iy_i^2x_{i1}^2 & \hdots & \sum_iy_i^2x_{i1}x_{i(p-1)}\\
        \vdots & \vdots & \ddots & \vdots\\
        \sum_iy_i^2x_{i(p-1)} & \sum_iy_i^2x_{i(p-1)}x_{i1} & \hdots & \sum_iy_i^2x_{i(p-1)}^2
            \end{bmatrix},
        \end{align*} 
        where we find that availability of the third and fourth joint sample moments involving the response variable makes it possible for this matrix to be computed. \\ 
        
        Similarly, the second term $2(y_i\mathbf{x}_i^T\boldsymbol{\beta})\mathbf{x}_i\mathbf{x}_i^T$ is also the product of a scalar $2(y_i\mathbf{x}_i^T\boldsymbol{\beta})$ and the matrix $\mathbf{x}_i\mathbf{x}_i^T$ resulting in
        \begin{align*}
            \begin{bmatrix}
        2(y_i\mathbf{x}_i^T\boldsymbol{\beta}) & 2(y_i\mathbf{x}_i^T\boldsymbol{\beta})x_{i1} & \hdots & 2(y_i\mathbf{x}_i^T\boldsymbol{\beta})x_{i(p-1)} \\
        2(y_i\mathbf{x}_i^T\boldsymbol{\beta})x_{i1} & 2(y_i\mathbf{x}_i^T\boldsymbol{\beta})x_{i1}^2 & \hdots & 2(y_i\mathbf{x}_i^T\boldsymbol{\beta})x_{i1}x_{i(p-1)}\\
        \vdots & \vdots & \ddots & \vdots\\
        2(y_i\mathbf{x}_i^T\boldsymbol{\beta})x_{i(p-1)} & 2(y_i\mathbf{x}_i^T\boldsymbol{\beta})x_{i(p-1)}x_{i1} & \hdots & 2(y_i\mathbf{x}_i^T\boldsymbol{\beta})x_{i(p-1)}^2 
            \end{bmatrix}.
        \end{align*}
        Summing over all observations leads to the matrix
        \begin{align*}
            \begin{bmatrix}
        2\sum_iy_i\mathbf{x}_i^T\boldsymbol{\beta} & 2\sum_ix_{i1}y_i\mathbf{x}_i^T\boldsymbol{\beta} & \hdots & 2\sum_ix_{i(p-1)}y_i\mathbf{x}_i^T\boldsymbol{\beta} \\
        2\sum_ix_{i1}y_i\mathbf{x}_i^T\boldsymbol{\beta} & 2\sum_ix_{i1}^2y_i\mathbf{x}_i^T\boldsymbol{\beta} & \hdots & 2\sum_ix_{i1}x_{i(p-1)}y_i\mathbf{x}_i^T\boldsymbol{\beta}\\
        \vdots & \vdots & \ddots & \vdots\\
        2\sum_ix_{i(p-1)}y_i\mathbf{x}_i^T\boldsymbol{\beta} & 2\sum_ix_{i(p-1)}x_{i1}y_i\mathbf{x}_i^T\boldsymbol{\beta} & \hdots & 2\sum_ix_{i(p-1)}^2y_i\mathbf{x}_i^T\boldsymbol{\beta} 
            \end{bmatrix}
        \end{align*}
        which can be computed from the third and fourth joint sample moments.\\
        
        Likewise, the final term $(\boldsymbol{\beta}^T\mathbf{x}_i\mathbf{x}_i^T\boldsymbol{\beta})\mathbf{x}_i\mathbf{x}_i^T$ summed over all observations becomes
        \begin{align*}
            \begin{bmatrix}
        \sum_i\boldsymbol{\beta}^T\mathbf{x}_i\mathbf{x}_i^T\boldsymbol{\beta} & \sum_i\boldsymbol{\beta}^Tx_{i1}\mathbf{x}_i\mathbf{x}_i^T\boldsymbol{\beta} & \hdots & \sum_i\boldsymbol{\beta}^Tx_{i(p-1)}\mathbf{x}_i\mathbf{x}_i^T\boldsymbol{\beta} \\
        \sum_i\boldsymbol{\beta}^Tx_{i1}\mathbf{x}_i\mathbf{x}_i^T\boldsymbol{\beta} & \sum_i\boldsymbol{\beta}^Tx_{i1}^2\mathbf{x}_i\mathbf{x}_i^T\boldsymbol{\beta} & \hdots & \sum_i\boldsymbol{\beta}^Tx_{i1}x_{i(p-1)}\mathbf{x}_i\mathbf{x}_i^T\boldsymbol{\beta}\\
        \vdots & \vdots & \ddots & \vdots\\
        \sum_i\boldsymbol{\beta}^Tx_{i(p-1)}\mathbf{x}_i\mathbf{x}_i^T\boldsymbol{\beta} & \sum_i\boldsymbol{\beta}^Tx_{i(p-1)}x_{i1}\mathbf{x}_i\mathbf{x}_i^T\boldsymbol{\beta} & \hdots & \sum_i\boldsymbol{\beta}^Tx_{i(p-1)}^2\mathbf{x}_i\mathbf{x}_i^T\boldsymbol{\beta} 
            \end{bmatrix}
        \end{align*}
        and is also composed of the third and fourth joint sample moments.

\end{document}